\documentstyle[12pt]{article}
\begin{document}
\begin{center}{\Large{\bf Pauli Spin Paramagnetism and Electronic Specific Heat in generalised $d$-dimensions}}
\end{center}

\vskip 1cm

\begin{center}{\it Muktish Acharyya}\\
{\it Department of Physics}\\
{\it Presidency University, 86/1 College Street,}\\
{\it Calcutta-700073, India}\\
{\it E-mail:muktish.acharyya@gmail.com}\end{center}

\vskip 1.5cm

\noindent {\bf Abstract:} The temperature variations of
pauli spin paramagnetic susceptibility
and the electronic specific heat of solids, are calculated as functions
of temperature following the free electron approximation, in generalised
$d$-dimensions. 
The results are compared and become consistent
with that obtained in three dimensions. Interestingly, the Pauli
spin paramagnetic susceptibility becomes independent of temperature
only in {\it two} dimensions.

\vskip 2cm


\newpage

\noindent {\bf Introduction:}

\vskip 1cm

As a consequence of having the intrinsic spin, the
electrons bear magnetic dipole moment. 
Pauli spin paramagnetism results entirely from the spin magnetic
moments of electrons.
In {\it three} dimensions, using free electron approximation, the
temperature dependent susceptibility is calculated by Stoner \cite{stoner}
and in the low temperature regime, it is quadratic in the temperature.
Following the same type of calculations,the electronic specific 
heat becomes linear in temperature. These results are well known and 
documented in the standard textbooks
\cite{huang,pathria,reif,schroeder}. 

However, no such derivations are observed in the standard literature, in
any generalised $d$-dimensions. 
Although, in reality, we are mainly concerned with the systems in three
dimensions, one may extend these calculations in generalised $d$-dimensions
just for pure pedagogical or mathematical interest.

In this paper, the Pauli spin paramagnetic susceptibility and the electronic
specific heat are calculated in free electron approximation as a function
of temperature in generalised $d$-dimensions.

\vskip 1cm
\noindent {\bf Calculations in $d$-dimensions:}

\vskip 0.3cm
\noindent {\it 1. Pauli spin paramagnetism:}

\vskip 1.3cm
In a generalised $d$-dimensions,
if we consider a system (unit volume) of electron gas, each electron will
contribute $-\mu_B$ ($\mu_B$ is Bohr magneton) to the magnetisation density
if its spin is parallel to the applied magnetic field $H$, and $\mu_B$, if
antiparallel. Hence, if $N_{\pm}$ is the number of electrons per unit
volume with spin parallel(+) or antiparallel(-) to $H$, the magnetisation
density will be \cite{mermin}
\begin{equation}
M = -\mu_B(N_{+} - N_{-}).
\end{equation}

\noindent When the interactions of the electrons with the magnetic field
will be considered, then the only effect of the field is to shift the
energy of each electronic level by $\pm \mu_B H$, according to whether
the spin is parallel or antiparallel to $H$. This may be expressed
in terms of the density of electronic states for a given spin.
If $g_{\pm}(E) dE$ is the number of electrons in the specified spin per
unit volume in the energy range $E$ and $E+dE$, one may write
\begin{equation}
g_{\pm}(E) = g(E)/2, ~~~~~~~~~~~~~~({\rm for~~~} H = 0)
\end{equation}

\noindent where $g(E)$ is the ordinary density of states. In the application
of external magnetic field $H$, the density of parallel and antiparallel
spin states are

\begin{eqnarray}
g_{+}(E) = {{1} \over {2}}g(E-\mu_B H){\nonumber}\\
g_{-}(E) = {{1} \over {2}}g(E+\mu_B H).
\end{eqnarray}

So, for each spin species, the number of electrons will be,

\begin{equation}
N_{\pm} = {\Large \int} g_{\pm}(E) F(E) dE,
\end{equation}

\noindent where $F(E)$ is the Fermi-Dirac distribution function\cite{reif}

\begin{equation}
F(E) = {{1} \over {e^{\beta (E-\mu)}+1}}.
\end{equation}
\noindent where, $\mu$ is in general
temperature dependent chemical potential.

For small field $H$, one may write

\begin{equation}
g_{\pm}(E) = {{1} \over {2}}g(E \pm \mu_B H) \simeq {{1} \over {2}}g(E)
\pm {{1} \over {2}} \mu_B H g'(E)
\end{equation}

\noindent and consquently,

\begin{equation}
N_{\pm} = {{1} \over {2}} {\Large \int_0^{\infty}}g(E) F(E) dE \mp {{1} \over {2}}
\mu_B H {\Large \int} g'(E) F(E) dE.
\end{equation}

Hence, the magnetisation density becomes,

\begin{equation}
M = \mu_B^2 H {\Large \int_0^{\infty}} g'(E) F(E) dE.
\end{equation}

Now, the primary task is to evaluate the following integral \cite{reif}
\begin{equation}
I = {\Large \int_0^{\infty}} g'(E) F(E) dE
\end{equation}

\noindent where, the derivative $g'(E)$ has to be calculted for the
electronic density of states $g(E)$ in a generalised $d$-dimensions.
This type of integral was evaluated in general in \cite{reif}
(page-393).
So,
\begin{eqnarray}
I &=&{\Large \int_0^{\infty}}g'(E)F(E) dE \nonumber\\
&=&\sum_{m=0}^{\infty}I_m{{(kT)^m} \over {m!}}[
{{d^mg(E)} \over {dE^m}}]_{\mu}\nonumber\\
&=& {\Large \int_0^{\mu}} g'(E) dE 
+ {{\pi^2 k^2 T^2} \over 6}g"(\mu)
+\sum_{m=4}^{\infty}I_m{{(kT)^m} \over {m!}}[
{{d^mg(E)} \over {dE^m}}]_{\mu}
\end{eqnarray}
\noindent where, $m=0,2,4,6...$ etc and $I_m=\int_{-\infty}^{\infty}
{{x^m e^x} \over {(e^x +1)^2}} dx$.
 $g"(\mu)$ is second derivative of $g(E)$ evaluated at $E=\mu$.
It may be noted here that only the even-derivatives (i.e., 
${{d^2g(E)} \over {dE^2}}, 
{{d^4g(E)} \over {dE^4}}, 
{{d^6g(E)} \over {dE^6}}$ etc.) of $g(E)$ are present in $I$.
Here, $\mu$ is in general temperature dependent chemical potential. 
The quantity
$I$ may be written as, 
\begin{eqnarray}
I &=& {\Large \int_0^{\mu_0}} g'(E) dE + 
{\Large \int_{\mu_0}^{\mu}} g'(E) dE  
+ {{\pi^2 k^2 T^2} \over 6}g"(\mu)
+\sum_{m=4}^{\infty}I_m{{(kT)^m} \over {m!}}[
{{d^mg(E)} \over {dE^m}}]_{\mu}
{\nonumber}\\
&\simeq& g(\mu_0) + g'(\mu_0)(\mu-\mu_0)
+ {{\pi^2 k^2 T^2} \over 6}g"(\mu),
+\sum_{m=4}^{\infty}I_m{{(kT)^m} \over {m!}}[
{{d^mg(E)} \over {dE^m}}]_{\mu}
\end{eqnarray}
\noindent where the second integral term is approximately equal to 
$g'(\mu_0)(\mu-\mu_0)$. Here, $\mu_0$ denotes the Fermi
energy ($T=0$) of the electron gas. Now, the task is to evaluate
$(\mu-\mu_0)$. It is well known \cite{reif}
that
\begin{equation}
\mu-\mu_0=-{{\pi^2 k^2 T^2 g'(\mu_0)} \over {6 g(\mu_0)}}
\end{equation} 
\noindent if calculated using quadratic (upto $T^2$ term) approximation.

After simplification (from above two equations), one may
write
\begin{equation}
I=g(\mu_0){\Large [}1 - 
 {{\pi^2 k^2 T^2} \over 6}{{g'(\mu_0)^2} \over {g(\mu_0)^2}}
+ {{\pi^2 k^2 T^2} \over 6}{{g"(\mu_0)} \over {g(\mu_0)}}{\Large ]}
+\sum_{m=4}^{\infty}I_m{{(kT)^m} \over {m!}}[
{{d^mg(E)} \over {dE^m}}]_{\mu}
\end{equation}
\noindent Now, to calculate $I$, one has to calculate, $g(\mu_0)$,
$g'(\mu_0)$ and $g"(\mu_0)$ in generalised $d$-dimensions. The density
of electronic states $g(E)$, in $d$-dimensions, has been calculated very
{\it recently} \cite{ma}. In $d$-dimensions, $g(E)=CE^{{d-2} \over 2}$, where
$C$ is a mass dependent constant. Thus, $g(\mu_0)=C\mu_0^{{d-2} \over 2}$,
$g'(\mu_0)=C({{d-2} \over 2})\mu_0^{{d-4} \over 2}$,
and $g"(\mu_0)=C({{d-2} \over 2})({{d-4} \over 2})\mu_0^{{d-6} \over 2}$.
Considering, $N=\int_0^{\mu_0}g(E) dE$, the total number of electrons, in
$d$-dimensions, $g(\mu_0) = dN/(2\mu_0)$. Using these, $I$ becomes
\begin{equation}
I={{dN} \over {2\mu_0}}{\Large [}1-({{\pi^2 k^2 T^2} \over {24\mu_0^2}})
((d-2)^2-(d-2)(d-4)){\Large ]}
+\sum_{m=4}^{\infty}I_m{{(kT)^m} \over {m!}}[
{{d^mg(E)} \over {dE^m}}]_{\mu}
\end{equation}
\noindent It should be noted here that, in the 
summation of the expression of $I$ in above equation
due to the presence of all 
even-derivatives
of $g(E)$, all terms include a factor $(d-2)$
(since $g(E)=CE^{{d-2} \over {2}}$).  
The magnetisation density $M=\mu_B^2 H I$. Hence, the pauli 
paramagnetic
susceptibility $\chi_P = M/H = \mu_B^2 I$ is calculated here in generalised
$d$-dimensions. 
This general (in $d$- dimensions) calculation is not yet done and it is
not found in the literature of quantum statistical physics \cite{huang,
pathria,reif,schroeder}.

The Pauli spin paramagnetic susceptibility is independent of the 
temperature for two-dimensional electronic systems when the higher 
order terms are included. However, this is based on the factor that 
the chemical potential is calculated using quadratic approximation 
in eqn (12). Actually, the temperature-independent feature of Pauli 
spin paramagnetic susceptibility for $d=2$ is very general. See eqn 
(8), for $d=2$, the derivative of density of state equals zero. Hence, 
the temperature dependent part of the magnetization and then the 
Pauli spin paramagnetic susceptibility vanish completely. 

Reader may easily check that
in three dimensions (putting $d=3$ in the general expression of $I$),
$\chi_P$ becomes (in quadratic approximation), 
\begin{equation}
{{3N\mu_B^2} \over {2\mu_0}}{\Large [} 1 - {{\pi^2 k^2 T^2} \over
{12 \mu_0^2}} {\Large ],}
\end{equation}
\noindent which was obtained by Stoner \cite{stoner, dekker}. 

Here, it may be noted that, in two dimensions {\it only}, 
(putting $d=2$ in the
general expression), the suceptibility becomes independent of temperature. 
This was not predicted by Stoner \cite{stoner}.
This is an {\it interesting} result and not yet available in standard
literature. This is a very general result also. Even if one calculate
the susceptibility using all terms in the summation (in equation 14), 
due to the presence
of a common factor of $(d-2)$(in all $T$ dependent terms), 
the Pauli spin susceptibility will become
temperature independent in two dimensions.
\vskip 2.0cm

\noindent {\it 2. Electronic Specific heat:}

\vskip 0.5cm

Here, the total energy ${\bar E}$ is calculated as \cite{reif}
\begin{equation}
{\bar E} = E_0 + {{\pi^2 k^2 T^2} \over 3}g(\mu_0),
\end{equation}
\noindent $E_0$ is the electronic energy at $T=0$. In $d$-dimensions
$g(\mu_0) = dN/(2\mu_0)$. Using this form of $g(\mu_0)$ the electronic
specific heat in $d$-dimensions, becomes
\begin{equation}
C_v = {{\pi^2 k^2 N d} \over {6 \mu_0}}T.
\end{equation}
\noindent The above result will yield well known result \cite{reif,dekker}
 of electronic specific heat ($C_v = (\pi^2 k^2 N T)/{2\mu_0}$) 
in three dimensions, if one puts $d=3$ in the general expression
of $C_v$ written above.

\vskip 1cm

\noindent {\bf Concluding remarks:}

\vskip 0.3cm

In this paper, the pauli spin paramagnetic susceptibility and electronic
specific heat are calculated in generalised $d$-dimensions. The results
are compared with that obtained in three dimensions. The susceptibility
and the specific heat are calculated as functions of dimensionality
$d$. The well known results in three dimensions are restored just by
putting $d=3$ in there respective general expressions. {\it Interesingly}, 
the Pauli spin paramagnetic susceptibility becomes independent of 
temperature {\it only} in two dimensions, which was not predicted by 
Stoner\cite{stoner}. These general calculations are not available in
the standard literatures of quantum statistical mechanics.
Although, $d$-dimensions is not a realistic concept, this study is 
pedagogically interesting and this result has a significance
in modern technology. If
any electronic device requires the 
temperature independent spin-magnetism, one may
think it to prepare in two dimensions.

It may be noted here that the Pauli spin susceptibility becomes 
independent of temperature in two dimensions 
but the electronic specific heat
does not. A possible explanation may be stated as follows: the spin
dependent magnetism and hence the susceptibility has a 
nontrivial dependence
on the topological arrangement of the electrons. As a result
the planar arrangement (in two dimensions) of electrons does not
produce any temperature dependent susceptibility. However, the
temperature dependence of the electronic kinetic 
energy has a trivial dependence
on the dimensionality. Hence, the 
temperature dependence of electronic specific heat is always
linear in any dimensions.

\end{document}